\documentclass[amsmath,amssymb,amsfonts,aps,pre,superscriptaddress,bibnotes,showpacs,showkeys,longbibliography,10pt,twocolumn,floatfix]{revtex4-2}
\usepackage{graphicx}
\usepackage{mathtools}
\usepackage{amsmath,amssymb,amsfonts}
\usepackage{sidecap}
\usepackage[english]{babel}
\usepackage{xcolor}
\usepackage{lipsum}
\usepackage{xr-hyper}
\usepackage{hyperref}

\DeclareMathOperator{\csch}{csch}

\makeatletter
\newcommand*{\addFileDependency}[1]{
\typeout{(#1)}
\@addtofilelist{#1}
\IfFileExists{#1}{}{\typeout{No file #1.}}
}\makeatother

\newcommand*{\myexternaldocument}[1]{%
\externaldocument{#1}%
\addFileDependency{#1.tex}%
\addFileDependency{#1.aux}%
}

\myexternaldocument{supplemental_material}

\begin{document}
\graphicspath{ {Figures/} }

\title{Positional information trade-offs in boundary-driven reaction-diffusion systems}

\author{Jonas Berx}
\affiliation{Niels Bohr International Academy, Niels Bohr Institute, University of Copenhagen, Blegdamsvej 17, 2100 Copenhagen, Denmark}
\author{Prashant Singh}
\affiliation{Niels Bohr International Academy, Niels Bohr Institute, University of Copenhagen, Blegdamsvej 17, 2100 Copenhagen, Denmark}
\author{Karel Proesmans}
\affiliation{Niels Bohr International Academy, Niels Bohr Institute, University of Copenhagen, Blegdamsvej 17, 2100 Copenhagen, Denmark}
\date{\today}

\begin{abstract}
Individual components such as cells, particles, or agents within a larger system often require detailed understanding of their relative position to act accordingly, enabling the system as a whole to function in an organised and efficient manner. Through the concept of positional information, such components are able to specify their position in order to, \emph{e.g.}, create robust spatial patterns or coordinate specific functionality. Such complex behaviour generally occurs far from thermodynamic equilibrium and thus requires the dissipation of free energy to sustain functionality. We show that in boundary-driven simple exclusion systems with position-dependent Langmuir kinetics, non-trivial Pareto-optimal trade-offs exist between the positional information, rescaled entropy production rate and global reaction current. Phase transitions in the optimal protocols that tune the densities of the system boundaries emerge as a result, showing that distinct protocols are able to exchange global optimality similar to phase coexistence in liquid-gas phase transitions, and that increasing the positional information can lead to diminishing returns when considering increased dissipation.

\end{abstract}

\maketitle

\section{Introduction}
One of the most intriguing processes in contemporary developmental biology is morphogenesis: how do patterns and functional structures such as tissues or limbs form from simple biochemical principles~\cite{bookMeinhardt}? Over the past century, two distinct mechanisms have been proposed to understand morphogenesis: Turing’s pattern formation in reaction-diffusion (RD) systems~\cite{Turingcite} and Wolpert’s concept of ‘positional information’ (PI)~\cite{WOLPERT19691}. Turing demonstrated that diffusing chemical species interacting via some activation-inhibition dynamics can, under appropriate conditions, spontaneously break the spatial symmetry and lead to the emergence of complex patterns like stripes and spots. On the other hand, in Wolpert's idea, the spatial symmetry inside a developing embryo is already broken due to the presence of a gradient of signalling molecules, aptly named morphogens. For instance, in the \emph{Drosophila} embryo, the locally deposited Bicoid (Bcd) morphogen at one end diffuses inside the embryo, establishing a concentration gradient~\cite{Gregordiffusion}. Cells then read out the local concentration and obtain information about their position. Hence, the morphogen signal is said to carry and transmit PI about cells. This is crucial for them to adopt fates that are appropriate for their location. Beyond biological systems, Wolpert's idea has also been realised using synthetic soft materials~\cite{Zadorin2017, BACCOUCHE2014234,toda2020engineering,dupin2022synthetic} where individual components such as bistable networks read out the morphogen gradient in a microchannel. This has facilitated improved controllability in experiments, making it suitable for testing new theoretical concepts.

Although PI remained an abstract concept for a few decades, a theoretical framework has recently been developed to quantify it~\cite{Dubuis2013, Gasper2014}. This information theory-based framework~\cite{Cover2006} defines PI as the mutual information between the position of the cell and other variables such as the gene expression levels or concentration of the signalling molecules, see eq.~\eqref{eq:mutual_info} below. With this, it has now become possible to measure PI even in experiments~\cite{Dubuis2013, Gregor2007, MRao2023, PRXLife.2.013016}. For example, four gap genes in the \emph{Drosophila} embryo have been found to provide approximately $\sim 4.2$ bits of PI, which enables cells to know their position with a precision of $\sim 1 \%$ of the total embryo length. Such a level of precision is critical for the development of robust body structures even though the surrounding environment is inherently noisy. We refer to Refs.~\cite{Gasper2021, Bialekbook2012} for a review on the PI framework.

Maintaining the gradients required for spatial patterning drives the system out of equilibrium. Given the importance of pattern formation, it is fundamental to understand the thermodynamic cost associated with RD and PI and non-equilibrium limits on their performance~\cite{falasco2018information, rana2020precision, Song2021, Tostevin2007,Emberly2008,Lo2015, singh2024}. From the perspective of stochastic thermodynamics, sustaining a required level of PI necessitates the dissipation of free energy to suppress biochemical noise in the morphogen gradients. This, in turn, can put fundamental limits on the precision with which position determination can be achieved~\cite{Tostevin2007,Emberly2008,Lo2015}. 

Over the past few decades, stochastic thermodynamics has proven to be a valuable framework for exploring these trade-offs~\cite{Seifert-review, VANDENBROECK20156,Seifert2005, TUR-1, TUR-2}. In this paper, we look at the PI through the lens of stochastic thermodynamics and explore the trade-off between PI, reaction current and the rescaled entropy production rate. To achieve this, we focus on the one dimensional boundary-driven simple symmetric exclusion process (SSEP) and augment it with position-dependent Langmuir kinetics, allowing for particle addition or removal in the bulk~\cite{PhysRevLett.87.150601, Derrida2002, Derrida_2007,Parmeggiani2004}. Such active transport of morphogens mediated by, \emph{e.g.}, kinesin or dynein motor proteins that propagate along neuronal axons has only recently been shown to control the axial polarity in, \emph{e.g.,} regenerating planaria, leading to a morphogen gradient carrying PI along the axons~\cite{Pietak2019,AGATA2014161,Li2018}.

Often in experiments, the morphogen profile is measured in one dimension: the anterior-posterior (AP) axis of the embryo~\cite{Dubuis2013,PRXLife.2.013016, Gregordiffusion} or the longitudinal axis in synthetic micro-channels~\cite{Zadorin2017}. Moreover, these signalling molecules experience a degradation effect inside the embryo~\cite{Grieneisen2012, Gregordiffusion}. This might occur due to enzymatic reactions taking place in the bulk, where a morphogen or substrate is transformed into a product molecule only if there is an enzyme present at its location~\cite{Buchner2013,Buchner2013_2}. This has led us to consider Langmuir kinetics in our model. Apart from this, the system is also connected with distinct particle reservoirs at its two ends, which drive it to a non-equilibrium steady state. We will focus on this steady state and illustrate the trade-off between PI and dissipation.

Furthermore, gradient formation is crucial not only for PI, but also for regulating particle fluxes that arise, \emph{e.g.,} due to reactions with localised hubs of DNA-bound Bcd~\cite{Ochoa-Espinosa2005,Mir2017}, or enzymatic complexes that form in order to catalyse multistep biochemical reactions~\cite{Heinrich1991,Hinzpeter2019,Hinzpeter2022}. Such complexes generally emerge from the spatial co-localisation (clustering) of multiple enzymes, drastically influencing the steady-state reaction flux as well as morphogen or substrate gradients~\cite{Buchner2013,Buchner2013_2}. For instance, carbon-fixing carboxysomes are specialised structures that compartmentalise enzymatic cascade reactions, allowing cyanobacteria to fix CO$_2$. Gradients of diffusible molecules such as CO$_2$ can create localised zones where carboxysome assembly is favourable~\cite{MacCready2020,MacCready2021}. These gradients provide PI by starting the carboxysome assembly in parts of the cytoplasm where substrate concentrations are higher and the flux associated with the carbon-fixing reactions can be optimised.


The paper is structured as follows. In sections~\ref{sec:Model} and~\ref{sec:PosInf} we respectively set up our boundary-driven system with Langmuir kinetics and define our observables of interest. Subsequently, in section~\ref{sec:Pareto}, we study the Pareto-optimal trade-offs between these observables for two different choices of the spatial distribution of Langmuir sites: clustered and uniform profiles, and briefly discuss how to compute approximate trade-offs for general Langmuir site distributions via a WKB approximation. Finally, in section~\ref{sec:conclusions}, we conclude and look toward future research avenues.

\begin{figure}
    \centering
    \includegraphics[width=0.95\linewidth]{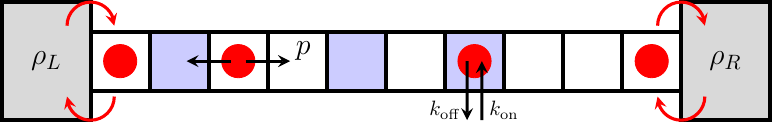}
    \caption{Schematics of the one dimensional lattice model. In the bulk, a particle can jump symmetrically to one of its neighbours with a constant rate $p$ if the target site is empty. Furthermore, a particle can attach to the $i$-th site with a site-dependent rate $k_{\rm on}^{i}$ if the site is unoccupied, and detach with rate $k_{\rm off}^{i}$; sites that allow such kinetics are shown by blue shading. At the boundaries, the system is connected to two particle reservoirs with average densities $\rho_L$ and $\rho_R$, that lead to particle addition or removal at these sites.}
   
 \label{fig:lattice}
\end{figure}
\section{Model}\label{sec:Model}
Let us set the stage by considering the one dimensional SSEP model which can be studied exactly~\cite{PhysRevLett.87.150601, Derrida2002, Derrida_2007,Parmeggiani2004}. It consists of $N$ lattice sites indexed as $i=1,\dots,N$, where $i=1$ and $i=N$ are the left and right boundaries, respectively, and the other sites are considered `bulk' sites, see Fig.~\ref{fig:lattice}. Within the bulk, a particle can jump symmetrically to one of its neighbouring sites with a constant rate $p$ if the target site is empty, obeying exclusion dynamics. Furthermore, a particle can attach to the $i$-th site with a site-dependent rate $k_{\rm on}^{i}$ if the site is unoccupied, and detach with a rate $k_{\rm off}^{i}$. The sites at which this can happen we henceforth call Langmuir sites. At the left and right boundaries, the system is connected to two particle reservoirs with average densities $\rho _L$ and $\rho _R$, respectively. Thus, particles can enter through these boundaries with rate $\alpha_{L/R}$ and exit with rate $\beta_{L/R}$, following the exclusion dynamics. Later, we will show how $\rho _L$ and $\rho _R$ are related to these rates. Due to coupling with these reservoirs, the system reaches a non-equilibrium steady state.

The occupation state of a site $i$ can be defined using a binary variable $n_i$ which can take two possible values, namely $n_i=1$ if the site is occupied and $n_i=0$ if it is empty. For the computation of PI, it is useful to calculate the average density $\rho _i (t)= \langle n_i \rangle $. To proceed with this calculation, we make the following choice for Langmuir rates \cite{Parmeggiani2004}:
\begin{align}
   k_{\rm on}^{i} = k_{\rm on}~e_i,~~~~ k_{\rm off}^{i} = k_{\rm off}~e_i\,,
\end{align}
where $k_{\rm on}$ and $k_{\rm off}$ are the rate constants independent of $i$. Here, we could imagine that particle attachment and detachment at the Langmuir sites are driven by the action of some enzyme $E$, with a local stationary concentration of $e_i$. We will set $k_{\rm B} = T = 1$ from now on. We adopt this viewpoint since there is an increasing interest in understanding the effects of enzyme arrangement on the performance of biological systems \cite{Heinrich1991,Hinzpeter2019,Hinzpeter2022, Parmeggiani2004}, although our results hold for any system with attachment-detachment kinetics. Here, we consider this to study its ramification on the PI.

One can write the dynamics of the average density $\rho _i (t)$ in the bulk site $i$ as
\begin{equation}
    \label{eq:time_avg_density}
    \begin{split}
        \dot{\rho_i} (t)&= p \left[ \rho_{i-1}(t) + \rho_{i+1}(t) -2 \rho_i(t)\right] + k_{\rm on} e_i   \\
        & ~~~~~- (k_{\rm on} + k_{\rm off}) e_i\rho_i(t)\, .
    \end{split}
\end{equation}
On the other hand, the density at the two boundary sites evolves as
\begin{align}
    & \dot{\rho_1} (t) = \alpha _L -(p+\alpha _L +\beta _L)\rho _1(t)+p\rho _2(t), \label{bcd-eq-1}\\
    & \dot{\rho_N} (t) = \alpha _R -(p+\alpha _R +\beta _R)\rho _N(t)+p\rho _{N-1}(t). \label{bcd-eq-2}
\end{align}
To simplify these equations, it is useful to introduce a new variable $\kappa = i/N$. While $i$ only takes discrete values, one can treat $\kappa \in [0,1]$ as a continuous variable for large $N$. In the remainder of our analysis, we will work in terms of this variable for mathematical simplicity since we can easily switch between $i$ and $\kappa$. We also denote $\rho _i(t)$ by $\rho (\kappa, t)$. For large $N$, one can perform the expansion
\begin{align}
    \rho_{i\pm 1}(t) \simeq \rho(\kappa, t)\pm \frac{1}{N} \frac{\partial  \rho(\kappa, t)}{\partial \kappa}+\frac{1}{N^2} \frac{\partial  ^2\rho(\kappa, t)}{\partial \kappa ^2} \label{ps-expansion}
\end{align}
and rewrite eq.~\eqref{eq:time_avg_density} as
\begin{equation}
    \label{eq:pde_full}
    \begin{split}
   \dot{\rho}(\kappa, t) &=\frac{p}{N^2}  \frac{\partial  ^2\rho}{\partial \kappa ^2} + k_{\rm on}\,e(\kappa)  - (k_{\rm on} + k_{\rm off})\,e(\kappa) \rho(\kappa, t)\,.
    \end{split}
\end{equation}
Let us next compare the effects of diffusion due to particle jumps and reaction kinetics. In absence of any Langmuir kinetics, the particle simply performs symmetric exclusion dynamics. To travel a distance of $\Delta i \sim N$ (or equivalently $\Delta \kappa \sim 1$), it will typically take a timescale $\Delta t_{\rm diff} \sim N^2$. Now, for the Langmuir kinetics to be on the same footing, the kinetic rates must scale as $1/N^2$. The typical timescale for these rates will then be $\Delta t_{\rm LK} \sim N^2$ and is of same order as $\Delta t_{\rm diff}$.
Therefore, we rescale the rates in eq.~\eqref{eq:pde_full} as $k_{\rm on} = \omega_{\rm on}/N^2$ and $k_{\rm off} = \omega_{\rm off}/N^2$ \cite{Parmeggiani2004}. The ratio $k_{\rm off}/k_{\rm on} = \omega_{\rm off}/\omega_{\rm on}$, however, remains unchanged in this scaling.

The resulting density equation in the steady state now becomes
\begin{equation}
    \label{ode_full}
    \rho''(\kappa) = \alpha^2 e(\kappa) \left(\rho(\kappa) - \gamma\right)\,,
\end{equation}
with $\gamma = \omega_{\rm on}/(\omega_{\rm on} + \omega_{\rm off})$ the Langmuir isotherm~\cite{Parmeggiani2004}, $\alpha^2 = (\omega_{\rm on} + \omega_{\rm off})/p$. The quantity $\alpha^2$ measures the competing effects of chemical reaction and diffusion and is generally referred to as the second Damk\"ohler number~\cite{Goppel2016,Almarcha2010}. Moreover, primes in eq.~\eqref{ode_full} indicate the derivatives with respect to $\kappa$. Repeating the same analysis in eqs.~\eqref{bcd-eq-1} and \eqref{bcd-eq-2} yields the average densities at the boundary reservoirs $\rho _L = \rho(0)$ and $\rho _R = \rho(1)$, with
\begin{equation}
\rho_{L} = \frac{\alpha_{L}}{\alpha_{L} + \beta_{L}}, ~\rho_{R} = \frac{\alpha_{R}}{\alpha_{R} + \beta_{R}}. \label{gaoinebx}
\end{equation}
When the reaction is driven strongly in the forward direction --where particles are absorbed by the bulk reservoir, the Langmuir isotherm $\gamma= 0$. In the reverse direction $\gamma = 1$ and at balance $\gamma = 1/2$, where $\omega_{\rm on} = \omega_{\rm off}$. Hence, we can use $\gamma$ as a measure of how far the system is from equilibrium, together with the difference in reservoir densities $\Delta\rho \equiv (\rho_L - \rho_R)$. 

Now that we have calculated the average density, it can be used to obtain the probability of the occupation number $n_i$ (or equivalently $n(\kappa)$). Making use of the binary characteristic of $n(\kappa)$, one can write
\begin{align}
    P(n|i=\kappa N) = \rho(\kappa) \delta _{n,1}+(1-\rho(\kappa)) \delta _{n,0}. \label{condd-ps}
\end{align}
This conditional probability is a key ingredient in the computation of PI, as we illustrate later. The central idea now is to solve the density equation~\eqref{ode_full} with boundary conditions $\rho(0) = \rho _L$ and $\rho(1) = \rho _R$ for a given distribution of Langmuir sites $e(\kappa)$. We then use it to obtain the conditional probability $P(n|\kappa)$ via eq.~\eqref{condd-ps} and finally to compute the PI. Subsequently, we will investigate optimal trade-offs between PI, the rescaled entropy production rate and the global reaction flux for different choices of $e(\kappa)$. Before that, let us calculate each of these quantities.

\section{Positional information and dissipation}\label{sec:PosInf}
\noindent\emph{Positional information --} In this section, we recall the mathematical framework introduced in Ref.~\cite{Dubuis2013} to quantify the PI. Quantifying information in a simple and unambiguous manner can be done by considering the mutual information $\mathcal{I}(X; Y)$ between two random variables $X$ and $Y$~\cite{Cover2006}. It is a measure of the dependence between $X$ and $Y$ and quantifies the reduction in uncertainty in one of the variables conditional on knowledge of another. In the context of PI, it provides a unique way to characterise the amount of information one variable (gene expression level, particle density, etc) provides about position~\cite{Dubuis2013,Jorswieck2014}.

Measuring the particle density in our lattice model, we gain a level of information about position equal to
\begin{equation}
    \label{eq:mutual_info}
    \mathcal{I} =  \sum _{n= \{0,1 \}}~\sum _{i=1}^{N} P(i,n) \log_2{\left(\frac{P(i,n)}{P_i(i) P_n(n)}\right)}\,,
\end{equation}
where $P(i,n)$ is the joint probability of position $i$ and occupation $n$, and $P_i(i),\,P_n(n)$ are the marginals. If we do not measure particle density on a site $i$, then we do not have any information about its position and it could anywhere inside the lattice. In probabilistic language, this means that the prior marginal distribution is uniform, $P_i(i) = 1/N$. This is also consistent with experiments on PI \cite{Dubuis2013, Gregor2007, MRao2023, PRXLife.2.013016, BACCOUCHE2014234}. Next, if we measure the particle density we gain information about the position of the site. For instance, if this site is found to be occupied, it is more likely to be closer to a particle source. There is a thus decrease in the uncertainty regarding the particle's position, and this reduction is the gained PI. The average information gained after measurement is represented by eq.~\eqref{eq:mutual_info}. Furthermore, $\mathcal{I}$ will be a function different model parameters. We have suppressed this dependence in the definition for notational simplicity.

Rewriting eq.~\eqref{eq:mutual_info} by means of Bayes' theorem
\begin{equation}
    \label{eq:mutual_info_rewrite}
    \mathcal{I} = \sum_{i=1}^{N}\, P_i(i) \left[S[P_n(n)] - S[P(n|i)]\right]\,, 
\end{equation}
one can express it as a difference between the Shannon entropy of the particle occupation probability and the Shannon entropy of the conditional probability~\cite{Cover2006}, \emph{i.e.,}
\begin{equation}
    \begin{split}
        S[P_n(n)] &= -\sum_{n=0,1} P_n(n) \log_2 P_n(n)\,, \\
        S[P(n|i)] &= -\sum_{n=0,1} P(n|i) \log_2 P(n|i)\,.
    \end{split}
\end{equation}
Recall that $P_n(n)  = \sum _{i=1}^{N} P(n|i)/N$ since the prior is uniform. Thus, one can calculate the PI simply from $P(n|i)$. Going back to our model, we see that this probability is given in eq.~\eqref{condd-ps}. Using this, we can thus compute the entropy associated with $P(n|i)$, \emph{i.e.},
\begin{equation}
    S[P(n|i)] = -\rho_i \log_2\rho_i - (1-\rho_i)\log_2(1-\rho_i)\,.
\end{equation}
Averaging over $P_i(i) = 1/N$ gives
\begin{align}
     & \sum _{i=1}^{N} P_i(i)~S[P(n|i)], \nonumber \\
    &~~~~=- \frac{1}{N} \sum _{i=1}^{N} \Big[\rho_i \log_2\rho_i + (1-\rho_i)\log_2(1-\rho_i)\Big]  \nonumber \\
   &~~~~ \simeq-\int_0^1 \mathrm{d} \kappa \Big[\rho(\kappa) \log_2\rho(\kappa) + (1-\rho(\kappa))\log_2(1-\rho(\kappa))\Big]\,. \label{new-ps-16-eq-1}
\end{align}
Where in the last line we have rewritten the expression with rescaled variable $\kappa = i/N$ and replaced $1/N~\sum _{i=1}^{N} \to \int _0 ^{1}~\rm{d} \kappa$ for large $N$. The approximate equality `$\simeq$' is used to indicate that it is valid only in the large $N$ limit.

Shifting now to the second term in the definition of $\mathcal{I}$ in eq.~\eqref{eq:mutual_info_rewrite}, we compute
\begin{align}
    P_n(n)  =& \frac{1}{N}\sum _{i=1}^{N} P(n|i) = \bar\rho ~\delta_{n,1} + (1-\bar\rho)~\delta_{n,0}\,, \label{eq:Pn_marginal} \\
    & \text{with }\bar{\rho} = \frac{1}{N} \sum _{i=1}^{N}\rho _i \simeq \int _0^{1} \rm{d} \kappa  ~\rho(\kappa). \label{hayvqp}
\end{align}
We then write the Shannon entropy associated with $P_n(n)$:
\begin{equation}
    \label{eq:entropy_average_density}
    S[P_n(n)] = -\left(\bar\rho \log_2\bar\rho + (1-\bar\rho)\log_2(1-\bar\rho)\right)\,,
\end{equation}
such that we can compute the PI through eq.~\eqref{eq:mutual_info_rewrite}, \emph{i.e.},
\begin{align}
        \mathcal{I} &= \int_0^1 \mathrm{d} \kappa \,\left[\rho( \kappa ) \log_2\rho( \kappa ) + (1-\rho( \kappa ))\log_2(1-\rho( \kappa ))\right] \nonumber \\
        &~~~~~~~~~~~~ - \left[\bar\rho\log_2\bar\rho + (1-\bar\rho)\log_2(1-\bar\rho)\right]\,. \label{eq:positional_information_full}
\end{align}
Importantly, note that the amount of PI that can be extracted from the system is symmetric around chemical equilibrium, \emph{i.e,} around $\gamma=1/2$. To see this, consider equation~\eqref{ode_full} for $\gamma\rightarrow 1-\gamma$ and introduce the density $\psi(x) = 1-\rho(x)$ of `holes' in the system. Plugging this into the above differential equation, we get
\begin{equation}
    \psi''(x) = \alpha^2 e(x) (\psi(x) - \gamma)\,,
\end{equation}
which is exactly equation~\eqref{ode_full}. Hence, the PI remains invariant under the substitution $\gamma\leftrightarrow 1-\gamma$, given judicious rescaling of the boundary values $\Delta\rho \leftrightarrow -\Delta\rho$.\\

\noindent\emph{Entropy production rate --} In our model, the system can be driven out of equilibrium due to contact with three particle reservoirs. First, the presence of two distinct boundary reservoirs causes the jumps between different lattice sites to break the detailed balance condition. Second, Langmuir kinetics occurring in the bulk can also violate this condition driving the system out of equilibrium. The total entropy production $\Delta S_{\rm tot}$ is equal to the entropy flux to each of the reservoirs, \emph{i.e.}, $\Delta S_{\rm tot} = \Delta S_{\rm res} = \sum_i Q^{(i)}/T^{(i)}$, with $Q^{(i)}$ the heat flowing to the $i$th reservoir at temperature $T^{(i)}$. We assume that $T^{(i)} =1$ and that the heat can be written as $Q^{(i)} = j^{(i)}\mu^{(i)}$, where $j^{(i)}$ is the particle flux and $\mu^{(i)}$ is the chemical potential or thermodynamic affinity, which can be determined by assuming local detailed balance~\cite{VANDENBROECK20156},
\begin{equation}
    \label{eq:ldb}
    \mu^{(L)} = \ln{\frac{\alpha_L}{\beta_L}}\,, \qquad \mu^{(R)} = \ln{\frac{\alpha_R}{\beta_R}}\,, \qquad \mu^{(B)} = \ln{\frac{k_{\rm off}}{k_{\rm on}}}\,,
\end{equation}
where the superscripts $L,R,B$ denote respectively the left, right and bulk reservoirs. 

Hence, the entropy production can be written as the sum of the individual contributions of each of the three reservoirs,
\begin{equation}
    \label{eq:entropy_production}
    \begin{split}
    \Delta S_{\rm tot} &\simeq N \int_0^1 \mathrm{d}\kappa \left[k_{\rm off} \rho(\kappa) - k_{\rm on} (1-\rho(\kappa))\right] e(\kappa) \ln{\frac{k_{\rm off}}{k_{\rm on}}} \\ 
    &+ \frac{p}{N} \rho'(0) \ln{\frac{\alpha_L}{\beta_L}} + \frac{p}{N} \rho'(1) \ln{\frac{\beta_R}{\alpha_R}}\,,
    \end{split}
\end{equation}
which can be rewritten by using eq.~\eqref{gaoinebx} as
\begin{equation}
    \begin{split}
    \Delta S_{\rm tot} &\simeq \frac{N}{k_{\rm off}+k_{\rm on}} \int_0^1 \mathrm{d}\kappa \left[\rho(\kappa)-\gamma\right] e(\kappa) \ln{\left(\frac{1-\gamma}{\gamma}\right)} \\ 
    &+ \frac{p}{N} \left[\rho'(0) \ln{\left(\frac{1-\rho_L}{\rho_L}\right)} + \rho'(1) \ln{\left(\frac{\rho_R}{1-\rho_R}\right)}\right]\,.
    \end{split}
\end{equation}
The term in square brackets in the integral can be simplified by inserting eq.~\eqref{ode_full} and integrating. We rescale the resulting equation by the Langmuir kinetics and simplify, resulting in
\begin{equation}
    \label{eq:entropy_production_rescaled}
    \begin{split}
        \Sigma &\equiv \frac{N \Delta S_{\rm tot}}{\omega_{\rm off} + \omega_{\rm on}} \\
        &\simeq\frac{1}{\alpha^2} \left[\rho'(0) \ln{\left(\frac{\gamma (1-\rho_L)}{\rho_L (1-\gamma)}\right)}-\rho'(1) \ln{\left(\frac{\gamma (1-\rho_R)}{\rho_R (1-\gamma)}\right)}\right]\,.
    \end{split}
\end{equation}

We will use this expression along with the PI from eq.~\eqref{eq:positional_information_full} to explore their trade-offs for different enzyme profiles. \\

\noindent\emph{Reaction flux --} Next, we will look at the third and final key quantity in this paper: the reaction flux. As discussed in the introduction, this quantity can play a pivotal role in setting up the density gradient which in turn will affect the amount of PI. To study its effect, we examine the
global reaction flux defined as 
\begin{align}
J = \sum_{i=1}^{N} \left[k_{\rm off} \rho_i - k_{\rm on} (1-\rho_i)\right] e_i\,.\end{align}
Following the same steps as before and also rescaling the flux with system size $\mathcal{J} = J N/(\omega_{\rm on} + \omega_{\rm off})$, one can obtain a simplified expression as
\begin{equation}
    \label{eq:reaction_flux_rescaled}
    \mathcal{J}  \simeq \int_0^1 \mathrm{d} \kappa \,e(\kappa) \left[ \rho(\kappa)-\gamma \right] = \frac{\rho'(1)-\rho'(0)}{\alpha^2}\,.
\end{equation}

Given that Langmuir kinetics offer a natural rescaling for the global reaction flux, we opted to rescale both the reaction flux and entropy production by $(\omega_{\rm on} + \omega_{\rm off})$. Alternatively, rescaling both observables by the diffusive timescale is also possible, though it would only introduce a multiplicative constant.

\section{Pareto-optimal trade-offs and phase transitions}\label{sec:Pareto}

We now consider the Pareto-optimal trade-offs between the PI $\mathcal{I}$, rescaled entropy production $\Sigma$ and the reaction current $\mathcal{J}$. Such trade-offs represent the mutual relation between these competing objectives~\cite{Miettinen2012}, indicating how the change in one objective impacts the others; it is the set of solutions to a multi-objective optimisation scheme where driving one objective closer to its optimal value negatively impacts the others. Since eq.~\eqref{ode_full} is not ubiquitously solvable for every choice of $e(\kappa)$, we will consider two examples for which exact analytical progress can be made and which closely follow related literature~\cite{Buchner2013,Buchner2013_2}: clustered and uniform profiles. While most of the trade-offs can be calculated exactly, we numerically determine other Pareto-optimal fronts by means of a high-precision genetic algorithm~\cite{Deb2008}, which in principle allows one to numerically study the Pareto fronts for any finitely supported choice of $e(\kappa)$.


\subsection{Clustered profile}
We choose the following sharply clustered profile $e(\kappa) = E_T\delta(\kappa-\kappa_0)$, with $0<\kappa_0<1$ and $E_T >0$. The reaction-diffusion equation~\eqref{ode_full} becomes the following:
\begin{equation}
\label{shaibe}
    \rho''(\kappa) = \alpha^2 (\rho(\kappa) - \gamma) \delta(\kappa-\kappa_0)\,,
\end{equation}
where we absorbed $E_T$ into the definition of $\alpha^2$. We solve the differential equation in the two regimes: $0\leq \kappa\leq \kappa_0$ and $\kappa_0\leq \kappa\leq1$. On both sides, the r.h.s. of the equation vanishes and the solution takes a linear form. Using the boundary conditions $\rho(0)=\rho_L$, $\rho(1)=\rho_R$ and requiring the continuity $\rho(\kappa \to \kappa _0^+) = \rho(\kappa \to \kappa _0^-) = \rho _0 $, we get
\begin{equation}
    \label{eq:density_profile_clustered}
    \rho(\kappa) = \begin{cases}
        \rho _L-(\rho _L-\rho _0)\frac{\kappa}{\kappa _0}, ~& 0\leq \kappa \leq \kappa_0 \\
        ~~\\
         \rho _R-(\rho _R-\rho _0)\frac{(1-\kappa)}{(1-\kappa _0)},~~~~ & \kappa_0 \leq \kappa \leq 1\,.
    \end{cases}
\end{equation}
To compute the density $ \rho _0$, we integrate eq.~\eqref{shaibe} from $-\epsilon$ to $\epsilon$ and take $\epsilon \to 0^+$. This gives
\begin{equation}
   \frac{d\rho}{d\kappa} \Bigg|_{\kappa \to \kappa _0^+}-\frac{d\rho}{d\kappa} \Bigg|_{\kappa \to \kappa _0^-} = \alpha^2 (\rho_0-\gamma)\,,
\end{equation}
Plugging in the above solution, we obtain
\begin{equation}
    \label{eq:rho_l/r}
    \rho _0 = \frac{(1-\kappa_0) (\alpha^2 \gamma \kappa_0 +\rho_L)+\kappa_0\rho_R}{A}
\end{equation}
with $A = 1+\alpha^2 \kappa_0 (1-\kappa_0)$. We now have all terms in the density $\rho(\kappa)$ in eq.~\eqref{eq:density_profile_clustered}. The PI $\mathcal{I}$ can be computed exactly by substituting $\rho(\kappa)$ in eq.~\eqref{eq:positional_information_full}. The exact expression is given in equation~\eqref{eq:PI_clustered_general} in section~\ref{app:app1} of the SM~\cite{supp}. From this expression, it is clear that the total PI is the sum of the contributions in the two domains separated by the delta function at $\kappa_0$, where the latter functions as a `reservoir' with particle density $\rho_0$, with an additional contribution from the coupling between the two domains. For $\alpha = 0$, the terms involving $\rho_0$ in SM eq.~\eqref{eq:PI_clustered_general} vanish and the PI reduces to the one derived in~\cite{singh2024}, where the resulting density profile is linear. Similarly, by setting $\kappa_0 = 0$ or $\kappa_0=1$, the reaction kinetics is pushed to the boundaries and the density profile also reduces to the one derived in~\cite{singh2024} and $\rho_0$ simply becomes either $\rho_L$ or $\rho_R$, respectively.

For this choice of $e(\kappa)$, there are five independent parameters: $\rho _L, \rho _R, \gamma, \alpha, \kappa _0$. We first use the analytic expression of $\mathcal{I}$ to carry out a multi-dimensional optimisation for these parameters. By using our genetic algorithm, we find that the 
the maximal PI is reached when the following conditions are met simultaneously: the distance from equilibrium is maximal, \emph{i.e.}, $\rho_L=1$, $\rho_R=0$ and $\gamma\downarrow 0$, and $\alpha\uparrow\infty$. Optimising next with respect to the location of the source, we find the optimal location is $x_0 = 1-\tanh{(1/2)}\approx 0.54$. Taking these limits in the expression for the PI, we find that the PI is bounded from above by
\begin{equation}
    \label{eq:I_max}
    \mathcal{I}_m \equiv \max_{\rho(x)}{\mathcal{I}} = \log_2\left(\frac{1+\mathrm{e}}{\mathrm{e}}\right)\approx 0.452\,.
\end{equation}
A detailed derivation of this result is given in section~\ref{app:app2} of the SM~\cite{supp}. Note that same result can be obtained by exchanging the boundary values and simultaneously setting either $x_0 = \tanh{(1/2)}$ or $\gamma=1$, due to the particle-hole symmetry; measuring the PI of the holes instead of the particles results in the same $\mathcal{I}_m$. Comparing this value with the one in the absence of Langmuir kinetics $(\alpha =0)$ where $\mathcal{I}_m =\log _2\left( 2/\sqrt{e}\right) \approx 0.278 $~\cite{singh2024}, we find that the single-site reaction kinetics enhance the maximum value of PI that can be conveyed.

Let us now understand the optimal parameters more heuristically. In this limit, the density $\rho_0$ becomes zero and the space can be segmented into two regions: the left region $0 \leq \kappa \leq \kappa _0$ and the right region $\kappa _0 \leq \kappa \leq 1$. In the left part, the maximum possible value of the density gradient is achieved when $\rho _L = 1$ for any given value of $\rho _R$. Now $\rho _R $ can vary anywhere between $[0,1]$. However any value of $\rho _R $ except $\rho _R  = 0 $ makes the overall density $\rho(\kappa)$ non-monotonic, which generally yields a smaller PI value than a monotonic profile. Only $\rho _R  = 0 $ gives a monotonic $\rho(\kappa)$, and hence large PI. Note that the maximum $\mathcal{I}_m$ requires $\rho _L=1$ and $\rho _R = 0$, which correspond to unidirectional transitions in the system~\eqref{gaoinebx}. Such a transition incurs infinite entropy production.

In some systems, it is important that the total number of morphogen particles that are reacted away is maximised, \emph{e.g.}, in order to deliver some product into the interior of the cell, before they are lost to the boundaries~\cite{Buchner2013,Buchner2013_2}, or to maximise the number of signalling events leading to pattern formation, \emph{e.g.,} in the binding of Bcd to patches of target genes~\cite{Ochoa-Espinosa2005,Mir2017}. The central object in this paper is therefore the trade-off between the bulk reaction flux, rescaled entropy production rate and the PI. In steady state, for the clustered profile the flux is given by
\begin{equation}
    \label{eq:bulk_flux_normalised}
    \mathcal{J} =  \rho_0 -\gamma\,,
\end{equation}
and the rescaled entropy production is given by
\begin{equation}
    \label{eq:entropy_production_delta}
    \begin{split}
        \Sigma &= \frac{1}{A\alpha^2}\left\{[\Delta\rho -\alpha^2 \kappa_0 (\rho_R -\gamma)] \ln{\left(\frac{\gamma (1-\rho_R)}{\rho_R (1-\gamma)}\right)}\right. \\
        &-\left.[\Delta\rho +\alpha^2 (1-\kappa_0) (\rho_L -\gamma)] \ln{\left(\frac{\gamma (1-\rho_L)}{\rho_L (1-\gamma)}\right)}\right\}\,.
    \end{split}
\end{equation}
Henceforth, we will assume that the reservoir densities $\rho_{L},\,\rho_{R}$ are the only tuneable parameters we have access to, since the values of $\alpha$, $\gamma$ and $x_0$ are generally fixed, depending on the specific chemical reaction or Langmuir kinetics, transport properties and clustering behaviour. In computing the Pareto-optimal trade-offs between $\mathcal{I},\,\mathcal{J}$ and $\Sigma$, we therefore look for optimal solutions within the set $(\rho_L,\rho_R)\in[0,1]^2$. \\

\begin{figure*}[htp]
    \centering
    \includegraphics[width=\linewidth]{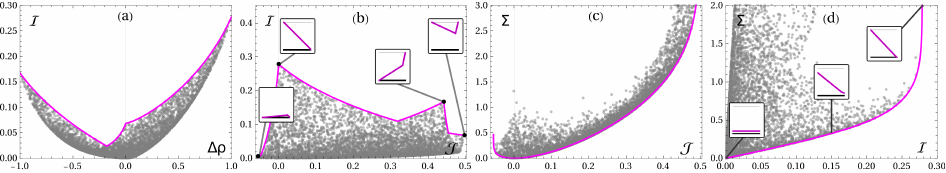}
    \caption{Trade-offs between $\mathcal{I},\,\mathcal{J},\,\Sigma$ and $\Delta\rho$ for a clustered profile $e(\kappa) = E_T \delta(\kappa-\kappa_0)$ with parameters $\alpha=3$, $\gamma=0.1$ and $\kappa_0 = 0.9$. Numerically generated results (grey dots) are obtained by uniformly drawing $(\rho_L,\,\rho_R)\in[0,1]^2$ and full lines are computed exactly in (a-c) or through numerical Pareto optimisation in (d). {\bf{(a)}} The upper bound on the PI as a function of $\Delta\rho$. Note that the upper bound exhibits kinks, where different solution branches intersect. {\bf (b)} Trade-off between the bulk reaction current $\mathcal{J}$ and the PI. The insets show the optimal density profiles corresponding to the local maxima (black circles) of the upper bound. These four points can be obtained by setting $(\rho_L,\,\rho_R)$ to either of the following: $\{(1,1),\, (1,0),\, (0,1),\, (0,0)\}$. {\bf (c)} Dissipation-current Pareto front, showing that $\Sigma$ diverges for $\mathcal{J} = -\gamma/A$ or $\mathcal{J} = (1-\gamma)/A$. {\bf (d)} Dissipation-PI Pareto front. The front is globally convex and the optimal density profiles (insets) transition smoothly from a constant $\rho(x)=\gamma$ to a monotonically decreasing piecewise profile.}
    \label{fig:plotgrid_clustered}
\end{figure*}

\indent
$\mathcal{I}-\Delta \rho$ \textit{trade-off} -- When varying the reservoir densities, we can plot $\mathcal{I}$ as a function of $\Delta\rho \equiv \rho_L - \rho_R$, see Fig.~\ref{fig:plotgrid_clustered}(a). The feasible combinations of both quantities are bounded from above and below. The lower bound suggests that as $\Delta\rho$ changes, a minimal amount of PI is unavoidably conveyed solely through changes in reservoir densities. This relationship is a monotonically decreasing (increasing) function of the density difference for $\Delta\rho\leq 0$ ($\Delta\rho \geq 0$). Since in this work we focus on maximisation of the PI, we will not consider this bound here. However, in the upper bound one can find `kinks', \emph{i.e,} discontinuities in the first derivative. These arise as a consequence of the intersection of two optimal solution branches as a function of $\Delta\rho$. One branch with $\rho_L=1$, $\rho_R = 1-\Delta\rho$ intersects another with $\rho_L = \Delta\rho$, $\rho_R=0$ for $\Delta\rho\geq 0$. Similarly, for $\Delta\rho\leq 0$, a branch with $\rho_L = 0$, $\rho_R = -\Delta\rho$ intersects another branch with $\rho_L = 1+\Delta\rho$, $\rho_R = 1$. In Fig.~\ref{fig:plotgrid_clustered}(a), a clear kink is visible for $\Delta\rho\leq 0$, while for $\Delta\rho\geq 0$, the corresponding kink is located very close to $\Delta\rho=1$ and is not visibile.

At small positive $\Delta \rho$, the density profile with $\rho_L = \Delta\rho$, $\rho_R=0$ is nearly flat and consequently the PI is small. So, the other optimal solution $\rho_L=1$, $\rho_R = 1-\Delta\rho$ dominates. However, as we increase $\Delta \rho$, the density profile becomes non-flat and its PI significantly increases. At some critical $\Delta \rho^*$, this solution dominates over the other one and a `kink' emerges in the upper bound. The critical density can be computed by equating the PI values of two optimal branches and solving it for $\Delta \rho$. Finally, our analytical study also gives the upper bound for $\Delta \rho \leq 0$, see Fig.~\ref{fig:plotgrid_clustered}(a). Here again, we find two possible optimal solutions, namely $\rho_L = 1+\Delta\rho$, $\rho_R = 1$ and $\rho_L = 0$, $\rho_R = -\Delta\rho$; one of them contributes to the upper bound depending on the value of $\Delta \rho$.\\


\indent
$\mathcal{J}-\mathcal{I}$ \textit{trade-off} -- For fixed values of $\alpha$, $\kappa_0$ and $\gamma$, the trade-off between the reaction flux and PI can exhibit multiple local maxima, corresponding to different possible optimal combinations of $\rho_L$ and $\rho_R$ that are able to simultaneously maximise both objectives, see the black circles in  Fig.~\ref{fig:plotgrid_clustered}(b). These four extremal points correspond to either $\Delta\rho = 0$, which leads to a generally lower PI as we showed, or to $\Delta\rho=1$. They can once again be determined by computing intersections between the aforementioned solution branches, where now the reaction flux can be computed as a function of $\Delta\rho$ and then parametrically drawn as a function of the PI, with $\Delta\rho$ taking values in $[-1,1]$. Interestingly, this upper bound is non-monotonic. This shows that while increasing the current generally entails a lowering of the PI, it can in fact be increased to a level at which it leads to an increasing PI. The `cusps' at local minima in the upper bound correspond to density profiles where $\rho_0$ is identical.

\indent
$\mathcal{J}-\Sigma$ \textit{trade-off} --We next investigate the relationship between the reaction current and the rescaled entropy production rate. Maximising $\mathcal{J}$ while simultaneously minimising $\Sigma$ leads to the exact Pareto-optimal trade-off between the two quantities. For a given $\mathcal{J}$, one can minimise $\Sigma$ by setting the two reservoir densities equal, $\rho_L = \rho_R = \rho$. This ensures that the entropy production due to the boundary drive is minimised to zero; the only contribution then arises from the particle flux at $\kappa = \kappa _0$. Plugging this in eq.~\eqref{eq:bulk_flux_normalised}, we can solve for $\rho(\mathcal{J})$ which we then substitute in eq.~\eqref{eq:entropy_production_delta}. This results in the following Pareto bound

\begin{equation}
    \label{eq:entropy_current_tradeoff}
    \Sigma(\mathcal{J}) = \mathcal{J} \left[\ln{\left(\frac{1-\gamma}{\gamma} \right)} + \ln{\left(\frac{\gamma + A \mathcal{J}}{1-\gamma-A\mathcal{J}}\right)}\right]\,.
\end{equation}
This bound is shown in Fig.~\ref{fig:plotgrid_clustered}(c). Note that $\Sigma(\mathcal{J})$ diverges when the reaction flux takes the values $\mathcal{J} = -\gamma/A$ or $\mathcal{J} = (1-\gamma)/A$ for $0<\gamma<1$. For these values of $\mathcal{J}$, the reservoir density is either $\rho = 0$ or  $\rho = 1$, which corresponds to unidirectional jumps into or out of the reservoirs in the model, see eq.~\eqref{gaoinebx}. This in turn amounts to a diverging entropy production rate $\Sigma$.\\

\indent
$ \Sigma-\mathcal{I}$ \textit{trade-off} -- The Pareto-optimal trade-off between the PI and rescaled entropy production, see Fig.~\ref{fig:plotgrid_clustered}(d), does not possess an obvious relation to $\Delta\rho$, and hence we cannot analytically determine it. However, it can be cast into a scalarised optimisation problem (a single-objective optimisation) with parameters $\rho_L,\,\rho_R$ using a linear combination of the objective functions,
\begin{equation}
    \label{eq:SOO}
    \Omega = -\lambda \mathcal{\mathcal{I}} + (1-\lambda) \Sigma\,,
\end{equation}
with $\lambda\in[0,1]$ a control parameter, such that any solution that minimises eq.~\eqref{eq:SOO} is on the Pareto optimal front. By tuning $\lambda$, we shift the focus of our optimisation protocol from the minimisation of the dissipation $(\lambda=0)$, to the maximisation of the PI $(\lambda=1)$.

Our numerically determined Pareto front is shown in Fig.~\ref{fig:plotgrid_clustered}(d). Smaller values of $\mathcal{I}$ suggest that there is no correlation between the position and the density. As such, the density profile associated with small $\mathcal{I}$ should be flat, with $\rho(\kappa) = \gamma $ (equilibrium). As a result, the rescaled entropy production rate is also equal to zero. As we increase $\mathcal{I}$, the density profile begins to become non-flat, approaching $\rho _L \to 1$ and $\rho _R \to 0 $ or vice versa. When $\mathcal{I}$ attains its maximum value, the front is characterised by a divergence of the entropy production. This is again due to the fact that the transitions from or into the reservoirs in the underlying model become unidirectional in this limit. This indicates that setting up a gradient that maximises the PI requires an increasing amount of dissipation. In between the two limits, the Pareto front is convex.

\begin{figure}[htp]
    \centering
    \includegraphics[width=0.8\linewidth]{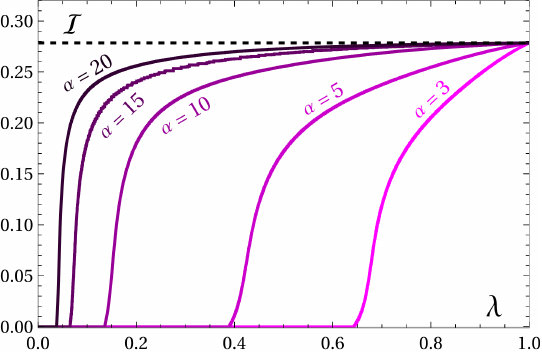}
    \caption{The PI as a function of $\lambda$ in the PI-rescaled entropy production Pareto front of the clustered profile. We have fixed $\gamma=0.1$ and $\kappa_0=0.9$ such that the maximal PI $\mathcal{I}_m = 1-1/\ln{4}$ (black dashed line) is independent of $\alpha$, showing that its influence is to decrease the critical value $\lambda_c$ with increasing $\alpha$. Since the PI is a continuous function of $\lambda$ with a discontinuity in the first derivative, the system undergoes a canonical second-order phase transition at $\lambda_c$.}
    \label{fig:phaseplot_clustered}
\end{figure}

Now, choosing the PI as an order parameter and sliding along the Pareto front in Fig.~\ref{fig:plotgrid_clustered}(d) by increasing $\lambda$, we see that every point on the front can be uniquely reached for a single value of $\lambda$, due to the convexity. However, $\mathcal{I} = 0$ for all $\lambda < \lambda_c$, where $\lambda_c \approx 0.643$ for the chosen parameters in Fig.~\ref{fig:plotgrid_clustered}. Increasing $\lambda$ further smoothly increases $\mathcal{I}$, showing that the transition from equilibrium where $\rho(x) = \gamma$ to non-equilibrium is of second order, see Fig.~\ref{fig:phaseplot_clustered}. The influence of increasing $\alpha$ is to decrease $\lambda_c$, and hence the slope of the Pareto front. Increasing $\alpha$ thus leads to a smaller dissipative cost per bit. 

Due to the convexity of the Pareto front, however, it becomes clear that a law of diminishing returns emerges~\cite{Tasnim2024,yadav2024,Balasubramanian2015,Tabbaa2014}, where a small increase in PI entails a significant increase of the associated cost.


\subsection{Uniform profile}
We now consider the second solvable model where the enzyme profile is a constant, \emph{i.e.}, $e(\kappa) = \bar{e}$. We absorb this value into the definition of $\alpha^2$ in eq.~\eqref{ode_full}, such that it becomes 
\begin{align}
\rho''(\kappa) = \alpha^2 \left(\rho(\kappa)-\gamma \right).
\end{align}
Solving it with boundary conditions $\rho _L$ and $\rho _R$ yields
\begin{equation}
    \label{eq:density_profile_uniform}
    \begin{split}
        \rho(\kappa) &= \gamma +\csch{(\kappa)}\left[(\rho_L-\gamma) \sinh{(\alpha (1-\kappa))} \right. \\
        &+ \left. (\rho_R-\gamma) \sinh{(\alpha \kappa)}\right]\,.
    \end{split}
\end{equation}
With this density profile, the PI associated with the system can be calculated using eq.~\eqref{eq:positional_information_full}. However, the results are long and complicated expressions that do not yield any particular insight; we list them in section~\ref{app:app1} of the SM~\cite{supp} for the case where $\gamma=0$. To find the maximum PI that can be conveyed, we first use again a genetic algorithm to gain intuition. We find that the optimal parameters are $|\Delta \rho| = 1$, $\gamma = 0,~\alpha = 0$, and the maximum PI is
\begin{align}
    \mathcal{I}_m = \log_2\left(2/\sqrt{e}  \right) \approx 0.278. \label{slau}
\end{align}
Next to PI, we also need the the reaction current and the rescaled entropy production rate. Similar to the clustered profile, these two quantities can be computed exactly as 
\begin{equation}
    \label{eq:current_uniform}
    \mathcal{J} = \bar{\rho}-\gamma\,,
\end{equation}
with $\bar{\rho}$ given in terms of $\rho(\kappa)$ in eq.~\eqref{hayvqp}, and
\begin{equation}
    \label{eq:entropy_production_uniform}
    \begin{split}
        \Sigma &= \frac{\csch{\alpha}}{\alpha}\left\{[(\rho_R-\gamma) - (\rho_L-\gamma) \cosh{\alpha}] \ln{\left(\frac{\gamma (1-\rho_L)}{\rho_L (1-\gamma)}\right)} \right. \\
        &+\left.[(\rho_L-\gamma) - (\rho_R-\gamma) \cosh{\alpha}] \ln{\left(\frac{\gamma (1-\rho_R)}{\rho_R (1-\gamma)}\right)}\right\}\,.
    \end{split}
\end{equation}\\

\begin{figure}[htp]
    \centering
    \includegraphics[width=0.95\linewidth]{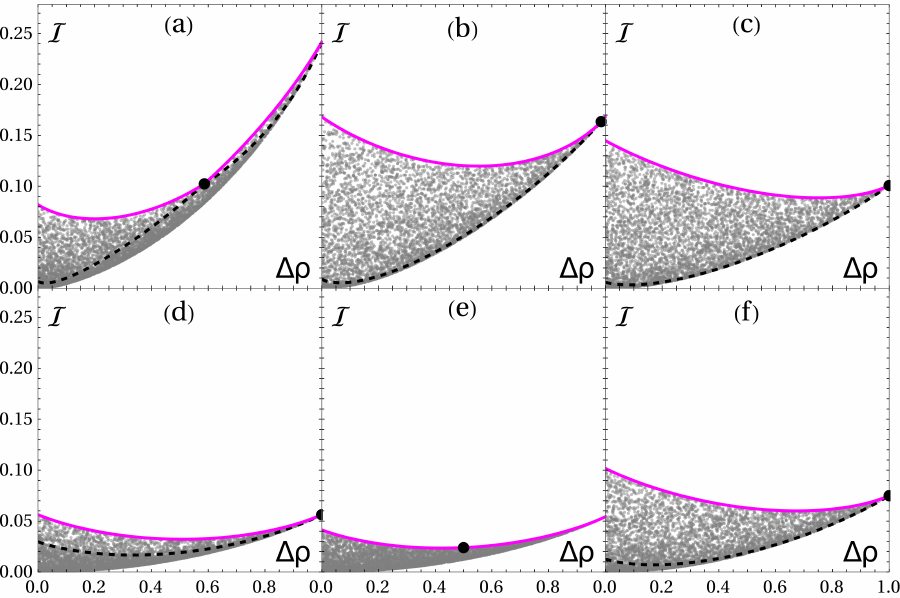}
    \caption{The PI $\mathcal{I}$ as a function of the reservoir density difference $\Delta\rho$ for a uniform $e(\kappa)$ profile. Panels {\bf (a-c)} show results for a fixed value of $\gamma=0.1$ while varying $\alpha$ as $\alpha = 3$, $\alpha = \alpha_c \approx 7.5$, and $\alpha = 15$, respectively. In panels {\bf (d-f)}, $\alpha$ is held constant at $\alpha = 15$, and $\gamma$ is varied with values $\gamma = \gamma_c \approx 0.4$, $\gamma = 0.5$, and $\gamma = 0.8$, respectively. Magenta lines indicate the upper bounds on the PI, with dashed black lines showing the analytical continuation on the bounds. Black circles indicate the point $\Delta\rho^*$ where bounds intersect.}
    \label{fig:igrid_constant}
\end{figure}

\indent $\mathcal{I}-\Delta \rho $ \textit{trade-off} -- Varying $\rho_L,\,\rho_R$, we plot the PI as a function of $\Delta\rho \equiv \rho_L - \rho_R$ in Fig.~\ref{fig:igrid_constant}. Due to the symmetry induced in the system by the uniform profile $e(\kappa)$, the density difference can be assumed to be positive, $\Delta\rho\geq0$ in this section.

Let us focus on the upper bound in Fig.~\ref{fig:igrid_constant}. The function that describes the upper bound on $\mathcal{I}$ is again given by two distinct curves that cross at a critical point $\Delta\rho^*$. These curves can be computed by solving the ODE~\eqref{ode_full} with the choices $\rho_L =1$, $\rho_R = 1-\Delta\rho$ and with $\rho_L = \Delta\rho$, $\rho_R = 0$. In Fig.~\ref{fig:igrid_constant}, the point $\Delta\rho^*$ is indicated by black circles. The upper bound on PI initially decreases as $\Delta\rho$ increases. This indicates that increasing the density difference between reservoirs reduces the system's PI. Beyond the minimum in the upper bound on the PI, however, increasing $\Delta\rho$ leads to an increase in the maximum PI. Thus, if the system aims to maximise PI while minimising the density difference, it is optimal to maintain $\Delta\rho = 0$ until it can be increased enough to achieve higher values of $\mathcal{I}$ at some $\Delta\rho>0$.

Furthermore, we observe that the upper bound reaches its maximum value at $\Delta \rho = 1$ for small $\alpha$ and at $\Delta \rho = 0$ for large $\alpha$, with a crossover occuring at some critical $\alpha _ c$. For low values of $\alpha$—characterised by small reaction rates, low enzyme concentrations, or high diffusion—the system can increase its PI by increasing the reservoir density difference, but this is only strictly true for $\alpha = 0$. When $\alpha > 0$ and for a fixed $\gamma$, reducing the reservoir density difference can, in fact, still enhance the PI with respect to intermediate values of $\Delta\rho$, see Fig.~\ref{fig:igrid_constant}. The maximum PI, however, is achieved when $\Delta\rho = 1$ and the density profile $\rho(\kappa)$ is a monotonic function.

\begin{figure}[htp]
    \centering
    \includegraphics[width=0.6\linewidth]{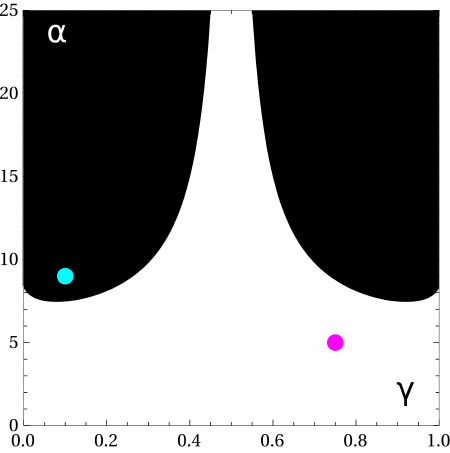}
    \caption{$(\gamma,\alpha)$ phase space for the uniform profile with the black regions showing the parameter combinations for which the maximal value of the PI can be achieved by setting the reservoir densities $\Delta\rho=0$, and the white region where $\Delta\rho=1$. The coloured circles indicate the parameter combinations we use for further analysis: $\gamma=0.75$, $\alpha=5$ (magenta), and $\gamma=0.1$, $\alpha=9$ (cyan).}
    \label{fig:phasespace_constant}
\end{figure}

\begin{figure*}[htp]
        \centering
        \includegraphics[width=0.95\linewidth]{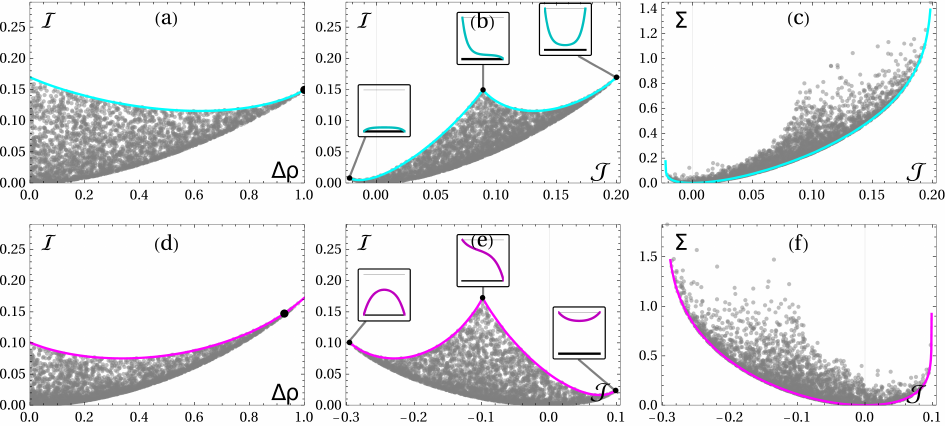}
        \caption{Trade-offs between $\mathcal{I},\,\mathcal{J},\,\Sigma$ and $\Delta\rho$ for a uniform profile. The top (bottom) row shows results for parameter combinations given by the cyan (magenta) circle in Fig.~\ref{fig:phasespace_constant}. {\bf (a,d)} The upper bound on the PI as a function of the density difference $\Delta\rho$. The black circle indicates $\Delta\rho^*$. {\bf (b,e)} Trade-off between PI and bulk current $\mathcal{J}$. The insets show that local maxima in the trade-offs correspond to either monotonic or non-monotonic morphogen density profiles. {\bf (c,f)} Dissipation-current trade-off~\eqref{eq:flux_entropy_tradeoff_uniform}, showing that $\Sigma$ diverges when the current is optimised~\eqref{eq:current_uniform_extrema}.
        Numerically generated results (grey dots) are obtained by uniformly drawing $(\rho_L,\,\rho_R)\in[0,1]^2$ and full lines are computed exactly.}
        \label{fig:plotgrid_uniform}
\end{figure*}

Conversely, for large $\alpha$, the density profile becomes flat in the bulk, with two peaks located at $\kappa=0$ and $\kappa=1$, making the density profile non-monotonic. In this scenario, the PI is primarily contributed by the edges, and it reaches its maximum when the two peaks are equal. This implies that for large $\alpha$, independent of the values of $\gamma$, the upper bound on PI is expected to be maximised when $\Delta\rho = 0$, which our calculations confirm. Thus, there is a crossover in the behaviour of the upper bound on PI as the parameter $\alpha$ is varied. One can also invoke a similar physical argument by varying $\gamma$ for a fixed value of $\alpha$. In Fig.~\ref{fig:phasespace_constant}, we have constructed exactly the $(\gamma,\alpha)$ phase space that shows two regions: a black region where the maximal value of the PI is achieved for $\Delta \rho = 0$ and a white one where it is attained for $\Delta \rho  =1$. The boundary delineating these two regions can be determined from the exact expression of the PI, as shown in Fig.~\ref{fig:phasespace_constant}. Here, the particle-hole symmetry is reflected in the phase space symmetry across the line $\gamma=1/2$. In Fig.~\ref{fig:plotgrid_uniform}(a,d), we show the $\mathcal{I}-\Delta\rho$ trade-off for both regions of the $(\gamma,\alpha)$ phase space. We now focus on the trade-offs between $\mathcal{I}$, $\mathcal{J}$ and $\Sigma$ and how they differ in both regions of the phase space.

\indent $\Sigma-\mathcal{J} $ \textit{trade-off} -- As in the clustered enzyme case, the relatively simple form of the global reaction flux allows us to determine the Pareto-optimal trade-off between the flux and the rescaled entropy production exactly, independent of whether the system is in the dark or light region of the $(\gamma,\alpha)$ phase space. Proceeding as before, we set $\rho_L = \rho_R = \rho$ and then solve for $\rho(\mathcal{J})$ in eq.~\eqref{eq:current_uniform}, yielding
\begin{equation}
    \rho\equiv \rho_L = \rho_R = \gamma + \frac{\alpha \mathcal{J}}{2}\coth{\frac{\alpha}{2}}\,.
\end{equation}
Plugging this into the density profile~\eqref{eq:density_profile_uniform} and calculating the rescaled entropy production~\eqref{eq:entropy_production_uniform}, we find that the optimal trade-off between rescaled entropy production $\Sigma(\mathcal{J})$ and flux $\mathcal{J}$ simplifies to
\begin{equation}
    \label{eq:flux_entropy_tradeoff_uniform}
    \Sigma(\mathcal{J}) = \mathcal{J}\left[\ln{\left(\frac{1-\gamma}{\gamma}\right)} + \ln{\left(\frac{2\gamma + \alpha\mathcal{J} \coth{\frac{\alpha}{2}}}{2(1-\gamma) - \alpha\mathcal{J} \coth{\frac{\alpha}{2}}}\right)}\right]\,.
\end{equation}
It can now be seen that the entropy production diverges when the current is optimised, \emph{i.e.}, when it reaches either of the following values
\begin{equation}
    \label{eq:current_uniform_extrema}
        \mathcal{J}_{\rm max} = \frac{2(1-\gamma)}{\alpha\coth{(\alpha/2)}} \,, \qquad \mathcal{J}_{\rm min} = -\frac{2\gamma}{\alpha\coth{(\alpha/2)}}\,,
\end{equation}
as shown in Fig.~\ref{fig:plotgrid_uniform}(c,f). Note that regardless of the value of $\gamma,\,\alpha$, the optimal currents always carry opposite sign.

\indent $\mathcal{I}-\mathcal{J} $ \textit{trade-off} -- Simultaneously optimising the PI and the current leads to a bound with three local PI maxima, see Fig.~\ref{fig:plotgrid_uniform}(b,e). Depending on whether the system is in the dark or light region of the phase space in Fig.~\ref{fig:phasespace_constant}, the maximal PI is achieved for either a non-monotonic or a monotonic density profile, respectively. However, while in the former case this maximum also yields the maximal current, solutions lying in the light region of phase space can only achieve maximal PI for negative values of the current, \emph{i.e.}, when there is a net inflow of morphogen particles into the bulk of the system. 

Taking a closer look at the local maximum of the PI as function of $\mathcal{J}$ corresponding to the monotonic density profile in Fig.~\ref{fig:plotgrid_uniform}(b), we see that it entails setting $\Delta\rho=1$, while the non-monotonic density profile corresponding to the global PI maximum is achieved by setting $\Delta\rho=0$. Both density profiles lead to a diverging entropy production~\eqref{eq:entropy_production_rescaled}. However, since they both lead to non-zero PI, we expect that the Pareto-optimal trade-off between PI and rescaled entropy production exhibits two optimal, diverging branches: one for monotonic and one for non-monotonic density profiles. 

\begin{figure}[htp]
    \centering
    \includegraphics[width=0.95\linewidth]{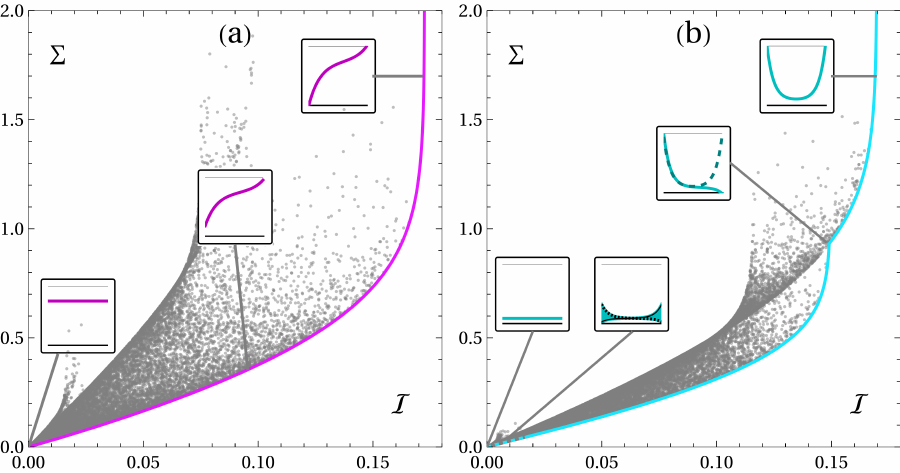}
    \caption{Pareto-optimal trade-offs between PI and rescaled entropy production for the parameter combinations indicated by coloured circles in Fig.~\ref{fig:phasespace_constant} for a uniform profile. In {\bf (a)}, the Pareto front (magenta) is fully convex while for {\bf (b)} the front (cyan) exhibits a concave region where the optimal density profile (insets) switches from a monotonic to a non-monotonic function, and a linear region (dotted) where multiple profiles coexist. Gray symbols are uniformly generated from $(\rho_L,\rho_R)\in[0,1]^2$. In the inset corresponding to the linear region in {\bf (b)}, we have drawn all coexisting solutions (cyan) together with the two bounding profiles (black dotted and full).}
    \label{fig:entropyIP_constant}
\end{figure}

\begin{figure}[htp]
    \centering
    \includegraphics[width=\linewidth]{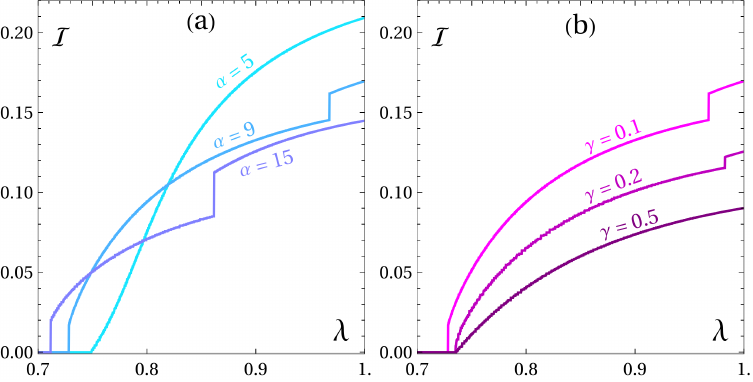}
    \caption{Phase transitions of the PI as a function of the tuning parameter $\lambda$ in the optimal protocol for the PI-rescaled entropy production trade-off or a uniform profile. In {\bf (a)} we fix $\gamma=0.1$ while varying $\alpha$ and in {\bf (b)} we fix $\alpha=9$ and vary $\gamma$. It is clear that in the dark regions in Fig.~\ref{fig:phasespace_constant} (cyan), the system exhibits a hybrid phase transition that carries signatures of both first and critical transitions, as well as a first order transition as a function of $\lambda$. Conversely, in the light region the system exhibits a single canonical second-order phase transition.}
    \label{fig:phaseplot_uniform}
\end{figure}

\indent $\mathcal{I}-\Sigma $ \textit{trade-off} -- Performing a scalar optimisation of the form~\eqref{eq:SOO} in the light region of the phase space shown in Fig.~\ref{fig:phasespace_constant} (magenta circle), now between $\mathcal{I}$ and $\Sigma$, we find that the optimal density profiles smoothly transition from a constant profile at $\lambda=0$ where $\rho(\kappa) = \gamma$ to monotonic profiles that maximize $\Delta\rho$ at $\lambda=1$, leading to a diverging entropy production at the maximal value of the PI, shown in Fig.~\ref{fig:entropyIP_constant}(a). Such a convex Pareto front once again leads to the type of second-order phase transitions we encountered for clustered profiles, see Fig.~\ref{fig:phaseplot_uniform}.

In the dark region of the phase space in Fig.~\ref{fig:phasespace_constant} (cyan circle), however, we find that at the critical value $\lambda_{c,1} \approx 0.728$ the system exhibits a non-trivial critical point, see Fig.~\ref{fig:phaseplot_uniform}, where signatures from both first and second order phase transitions can be found~\cite{Mora2011,Tkačik2015,seoane2016}. For $\lambda<\lambda_{c,1}$, the optimal profile is simply constant, $\rho(\kappa) = \gamma$. At criticality, however, multiple optimal monotonic density profiles coexist, since the Pareto front corresponding to the value of $\lambda=\lambda_{c,1}$ is linear, see the inset corresponding to the dashed part of the front in Fig.~\ref{fig:entropyIP_constant}(b). Increasing $\lambda$, the optimal profiles increase PI by increasing $\Delta\rho$, with a higher associated energy cost. 

At a second critical value $\lambda_{c,2} \approx 0.967$, the Pareto front becomes locally concave and a first order phase transition occurs. The earlier monotonic optimal density profiles now exchange global optimality with non-monotonic profiles where the system can increase PI by lowering $\Delta\rho$. Such local non-convexity of the Pareto front between the mutual information and entropy production has recently also been observed for general communication channels~\cite{Tasnim2024}. For $\lambda>\lambda_{c,2}$, the optimal profiles remain non-monotonic and PI can be further increased only by expending progressively larger amounts of free energy. The exchange of global optimality shows that the system can quickly switch between protocols that optimise dissipation to ones that optimise PI.

In the concave region of the Pareto front, the optimal density profiles lose monotonicity at the cusp for a PI value that can be computed by setting $\rho_L = 1,\,\rho_R=0$ (or vice versa); the solution is shown in eq.~\eqref{eq:Ic} in section~\ref{app:app1} of the SM~\cite{supp}. These `metastable' profiles can be accessed only by considering hysteretic protocols, where varying the values of $\rho_L,\,\rho_R$ can trap the system in locally stable states. This region represents a thermodynamically suboptimal zone for communication.


\subsection{General profiles}
So far, we have studied the trade-offs for two solvable models. We will now use the gained intuition to approach the calculations and trade-offs for general choices of $e(\kappa)$. There is no general way to solve eq.~\eqref{ode_full} for any choice of Langmuir distribution profile $e(\kappa)$. However, when $e(\kappa)$ varies slowly with respect to the characteristic decay length of the density, a small-noise (WKB) approximation can be used to find approximate density profiles. Making the change of variables $\psi(\kappa) = \rho(\kappa) - \gamma$, with corresponding boundary values $\psi_{L/R} = \rho_{L/R}-\gamma$, the equation for the morphogen density~\eqref{ode_full} becomes 
\begin{equation}
    \label{eq:ode_reduced}
    \frac{1}{\alpha^2} \psi''(\kappa) = e(\kappa)\,\psi(\kappa)\,.
\end{equation}
Setting now $\epsilon = 1/\alpha$, eq.~\eqref{eq:ode_reduced} reduces to a well-known simple form for the WKB approximation~\cite{Bender2010}. This technique is valid for small values of $\epsilon \ll 1$, so we will assume that the system is in a regime where the Langmuir kinetics dominate, \emph{i.e.}, $\alpha \gg 1$. We assume then a solution of the form $\psi(x)\sim \exp{\frac{1}{\delta}\sum_{n=0}^\infty \delta^n S_n(x)}$. Inserting this into~\eqref{eq:ode_reduced}, we find that $\delta=\epsilon$ by dominant balance. Collecting terms in powers of $\epsilon$ up to $\mathcal{O}(\epsilon)$, we find that after applying the boundary conditions, the first-order solution in the WKB approximation becomes
\begin{equation}
    \label{eq:solution_WKB_first}
    \begin{split}
    \psi(\kappa) &\sim \csch{\left(F(0,1)\right)} \left\{\psi_L \left(\frac{e_L}{e(\kappa)}\right)^\frac{1}{4}\sinh{\left(F(\kappa,1)\right)}\right. \\
    &+\left.\psi_R \left(\frac{e_R}{e(\kappa)}\right)^\frac{1}{4}\sinh{\left(F(0,\kappa)\right)}\right\}
    \end{split}
\end{equation}
with $F(s,t) = \alpha \int_s^t \mathrm{d}q\,\sqrt{e(q)}$, $e_L = e(0)$ and $e_R = e(1)$. The solution~\eqref{eq:solution_WKB_first} differs from the exact solution to eq.~\eqref{eq:ode_reduced} by terms of order $\alpha^{-1}$ for $e(\kappa)\neq 0$. For instance, setting $e(\kappa)=1$ recovers the exact density function~\eqref{eq:density_profile_uniform} for the uniform Langmuir profile. Conversely, the WKB approximation breaks down for rapidly varying $e(\kappa)$.

Performing the integrals involved in calculating the PI~\eqref{eq:positional_information_full} is generally analytically not possible, but they can be performed numerically. The rescaled entropy production~\eqref{eq:entropy_production_rescaled} and reaction current~\eqref{eq:reaction_flux_rescaled}, however, depend only on derivatives of the density profile at the boundaries, so they can be computed exactly; we list the expressions in section~\ref{app:app3} of the SM~\cite{supp}.

\begin{figure}[htp]
    \centering
    \includegraphics[width=0.85\linewidth]{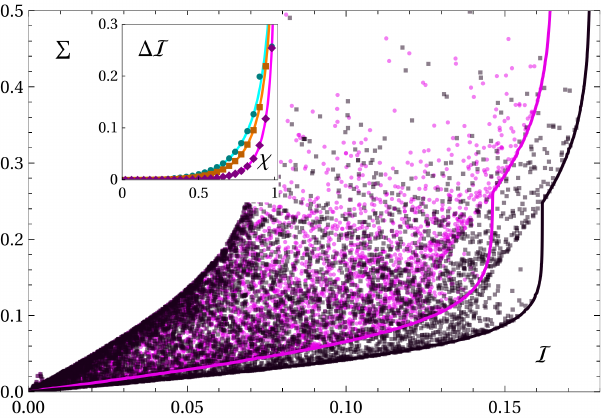}
    \caption{Pareto-optimal PI-dissipation trade-off for the exact (black) solution to eq.~\eqref{ode_full} versus the WKB approximation~\eqref{eq:solution_WKB_first} (magenta) for linear profiles $e(\kappa) = (1-\kappa) e_L + \kappa e_R$, with $\alpha=20$, $\gamma=0.1$ and $e_L = 0.05$. {\bf Inset:} fractional change $\Delta\mathcal{I}\equiv1 - \mathcal{I}_m^{\rm (wkb)}/\mathcal{I}_m^{(\rm ex)}$ as a function of the slope $\chi$, for $\alpha=20$ (magenta), $\alpha=10$ (orange) and $\alpha=5$ (cyan). Symbols are numerically computed through a genetic algorithm, lines are analytically determined.}
    \label{fig:WKB}
\end{figure}

To see how well the WKB approximation reproduces the exact results, we will choose a linear profile $e(\kappa) = (1-\kappa) e_L + \kappa e_R$, for which the exact solution to eq.~\eqref{ode_full} is a complicated expression involving Airy functions. Without loss of generality, we can set $e_R = 1-e_L$, since this simply entails rescaling $\alpha$. We can then compare how the approximate solutions hold up against the exact one on the level of the Pareto fronts by considering the PI-dissipation trade-off, shown in Fig.~\ref{fig:WKB}. 

We compare the WKB approximation to the exact solution by considering the fractional change of the difference of the maximal values of the PI, i.e, $\Delta\mathcal{I}\equiv1 - \mathcal{I}_m^{\rm (wkb)}/\mathcal{I}_m^{(\rm ex)}$, as a function of the slope $\chi = (e_R - e_L)/(e_R +e_L)$ of the Langmuir density profile, see the inset in Fig.~\ref{fig:WKB}. The WKB approximation becomes more accurate when the slope is small, since then $e(\kappa)$ varies more slowly with respect to $\rho(\kappa)$, and with increasing $\alpha$. There exists now again a region in the $(\alpha,\gamma,e_L)$ phase space where the PI-dissipation trade-off displays a concave region. We have checked numerically for some parameter values that this concavity is similarly induced by the switching of monotonic to non-monotonic density profiles in the optimal protocol.

\section{Conclusions}\label{sec:conclusions}
In this paper, we have integrated positional information -- a fundamental concept in developmental biology -- into the framework of stochastic thermodynamics. We introduced a simple model for communicating PI in boundary-driven exclusion processes with position-dependent Langmuir kinetics and analysed Pareto-optimal trade-offs among PI, rescaled entropy production, and global reaction current for clustered and uniform Langmuir site distributions. In the clustered distribution, maximizing the reservoir density difference leads to monotonic particle density profiles that achieve the highest PI but at the expense of increased rescaled entropy production. Conversely, when maximizing reaction current, the system optimises PI by alternating between monotonic and non-monotonic density profiles, with the latter maximising current when minimising the reservoir density difference. This trade-off between PI and rescaled entropy production is convex, suggesting a thermodynamic advantage in spreading information across multiple lower-capacity channels: inverse multiplexing. This concept has recently been hypothesized to exist in general communication systems that dissipate free energy to optimise channel capacity~\cite{Tasnim2024,yadav2024}. We leave the application of that idea to our model as an exciting future research avenue.

For uniform Langmuir distributions, the optimal protocol varies with system parameters, leading to two distinct regimes. In the first regime, optimal trade-offs resemble those of the clustered profile, with monotonic density profiles yielding the highest PI, while non-monotonic profiles maximise current but convey a smaller amount of PI. The optimal rescaled entropy production again forms a convex trade-off with PI. The second regime, however, features a concave section in the PI-dissipation Pareto front, indicating a thermodynamically suboptimal zone where the optimal density profile shifts from monotonic to non-monotonic with a small reservoir density difference. Next to these two exactly solvable choices of $e(\kappa)$, we also demonstrated that the WKB approximation can yield qualitatively accurate trade-offs for general Langmuir density profiles if they vary slowly on the solution's characteristic length scale.

Going beyond the simple dynamics of our toy model, it would be interesting to see how the optimal trade-offs can be modified with the inclusion of active driving such as the asymmetric simple exclusion process (ASEP), which is used to model directional transport of \emph{e.g.}, kinesin or dynein, involved in morphogenesis~\cite{Tischer2010}. We also believe that our theoretical study opens up new possible avenues for experiments such as designing cost-efficient PI strategies or probing the trade-offs and phase transitions. \\

\emph{Data availability statement --} All data that support the findings of this study are included within the article (and any supplementary files)

\begin{acknowledgements}
J.B. is funded by the European Union’s Horizon Europe framework under the Marie Sk\l odowska-Curie grant agreement no. 101104602 ‘STBR’. K.P. is funded by the European Union’s Horizon 2020 research and innovation program under the Marie Sk\l odowska-Curie grant agreement no. 101064626 ‘TSBC’, and from the Novo Nordisk Foundation (grant nos. NNF18SA0035142 and NNF21OC0071284).
\end{acknowledgements}

\bibliographystyle{apsrev4-2}
\bibliography{biblio.bib}

\end{document}


\graphicspath{ {Figures/} }

\title{Supplementary Material: Positional information trade-offs in boundary-driven reaction-diffusion systems}

\author{Jonas Berx}
\affiliation{Niels Bohr International Academy, Niels Bohr Institute, University of Copenhagen, Blegdamsvej 17, 2100 Copenhagen, Denmark}
\author{Prashant Singh}
\affiliation{Niels Bohr International Academy, Niels Bohr Institute, University of Copenhagen, Blegdamsvej 17, 2100 Copenhagen, Denmark}
\author{Karel Proesmans}
\affiliation{Niels Bohr International Academy, Niels Bohr Institute, University of Copenhagen, Blegdamsvej 17, 2100 Copenhagen, Denmark}
\date{\today}

\maketitle

\section{Exact expressions for positional information}
\label{app:app1}

\subsection{Clustered profile}
The positional information $\mathcal{I}$ can be computed exactly for the clustered profile and is given by the simplified expression

\begin{equation}
    \label{eq:PI_clustered_general}
    \mathcal{I} = \kappa _0 ~\mathcal{Y}(\rho _L, \rho _0)+ (1-\kappa _0)~ \mathcal{Y}(\rho _R, \rho _0) - \bar{\rho} \log _2 \bar{\rho}-(1-\bar{\rho}) \log_2 (1-\bar{\rho})
\end{equation}
where we use the notation
\begin{align}
&\bar{\rho} = \frac{\kappa _0}{2}(\rho _L+\rho _0)+\frac{(1-\kappa _0)}{2}(\rho _R+\rho _0) , \\
&\mathcal{Y}(a,b)  = \frac{1}{(a-b)\ln 4} \left[ a^2 \ln a-b^2 \ln b -(1-a)^2 \ln (1-a)+(1-b)^2 \ln (1-b)-(a-b)\right]
\end{align}

\subsection{Uniform profile}
The term involving the average Shannon entropy of the conditional expectation, \emph{i.e.,} $\langle S[P(n|\kappa)]\rangle$ for the case where $\gamma=0$ is given by the following expression,
\begin{widetext}
\begin{equation}
    \begin{split}
        -\langle S[P(n|\kappa)]\rangle \cdot \ln{2} &= 1+\frac{\pi^2}{6 \alpha} - \ln{\left(\frac{1-e^{2\alpha}}{1-e^\alpha (1-\Delta\rho)}\right)} + \frac{1}{\alpha}\coth{\frac{\alpha}{2}} \Delta\rho \log{\Delta\rho} 
        + \frac{2-\Delta\rho}{\alpha} \tanh{\frac{\alpha}{2}} - \frac{2 \mathcal{A}}{\alpha}\csch{\alpha} \ln{\frac{1}{\mathcal{A}}\cosh{\frac{\alpha}{2}}} \\
        &-\frac{1}{\alpha}\left(\Li_2(e^\alpha) - \Li_2\left(\frac{1-\Delta\rho-e^\alpha}{1-e^\alpha (1-\Delta\rho)}\right)+\Li_2\left(\frac{e^\alpha (1-\Delta\rho -e^\alpha)}{1-e^\alpha (1-\Delta\rho)}\right)\right) \\
        &- \frac{2\csch{\alpha}}{\alpha}\left\{\sinh^2{\frac{\alpha}{2}}\left[(2-\Delta\rho) + \ln{(1-\Delta\rho)}\left(\Delta\rho (1+\frac{1}{2}\csch^2{\frac{\alpha}{2}})-1\right)\right]\right.\\
        &+\left.\mathcal{B} \left(\arctan{\frac{\Delta\rho}{\mathcal{B}}} + \arctan{\frac{\mathcal{A}}{\mathcal{B}}}\right)\right\}\,,
    \end{split}
\end{equation}
\end{widetext}
where $\mathcal{A} = 1-\Delta\rho-\cosh{\alpha}$ and $\mathcal{B}^2 = 4 \sinh^2{\left(\frac{\alpha}{2}\right)} (1-\Delta\rho) - \Delta\rho^2$. The other term involving the entropy associated with $P(n)$, \emph{i.e.,} $\langle S[P(n)]\rangle$ is given by equation~\eqref{eq:entropy_average_density} in the main text, with the average density given by
\begin{equation}
    \bar{\rho} = \gamma + \frac{2(1-\gamma)-\Delta\rho}{\alpha}\tanh{\frac{\alpha}{2}}\,.
\end{equation}

This is only the upper bound on the PI for $\Delta\rho < \Delta\rho^*$. After the transition the shape of the bound changes. In this regime the term involving the average Shannon entropy of the conditional expectation, \emph{i.e.,} $\langle S[P(n|\kappa)]\rangle$ for the case where $\gamma=0$ is given by 
\begin{widetext}
\begin{equation}
    \begin{split}
        -\langle S[P(n|\kappa)]\rangle \cdot \ln{2} &= 1-\frac{\alpha}{2} + \frac{\csch{\alpha}}{\alpha} \left\{2 \mathcal{C} \arctanh{\frac{\Delta\rho}{\mathcal{C}}} -\Delta\rho \ln{\frac{2\Delta\rho}{\sinh{\alpha}}}
        +\sinh{\alpha} \left[\Li_2\left(\frac{\Delta\rho}{\sinh{\alpha}-\mathcal{C}}\right)+\Li_2\left(\frac{\Delta\rho}{\sinh{\alpha}+\mathcal{C}}\right)\right]\right\} \\ 
        &+ \frac{2\mathcal{C}\csch{\alpha}}{\alpha}\arctanh{\frac{\mathcal{A}}{\mathcal{C}}}-\frac{\Delta\rho}{\alpha}\coth{\alpha}\ln{\frac{1-\Delta\rho}{\Delta\rho}}
        + \frac{\Delta\rho}{\alpha}\csch{\alpha}\ln{\coth{\frac{\alpha}{2}}}+\ln{\frac{\Delta\rho}{2\sinh{\alpha}}} \\
        &- \frac{1}{\alpha}\left[\Li_2\left(\frac{e^\alpha \Delta\rho}{\sinh{\alpha}+\mathcal{C}}\right)+\Li_2\left(\frac{e^\alpha \Delta\rho}{\sinh{\alpha}-\mathcal{C}}\right)\right]\,,
    \end{split}
\end{equation}
\end{widetext}
with $\mathcal{C} = \Delta\rho^2 + \sinh^2{\alpha}$. The other term is then again given by equation~\eqref{eq:entropy_average_density} in the main text, with average density
\begin{equation}
    \bar{\rho} = \gamma + \frac{\Delta\rho -2\gamma}{\alpha} \tanh\frac{\alpha}{2}\,.
\end{equation}

The full expression with general values of $\gamma$ possesses a similar structure but is too lengthy to show here. However, the expressions simplify somewhat in the limit where $\Delta\rho = 1$, with the choice $\rho_L = 1,\, \rho_R = 0$. This then results in the exact expression for the critical value $\mathcal{I}_c$, where the $\Sigma-\mathcal{I}$ trade-off presents a cusp inside the locally concave region. It can be computed exactly as
\begin{equation}
    \label{eq:Ic}
    \begin{split}
        \mathcal{I}_c \ln{2} &= 1+\frac{\alpha\gamma}{2} + \gamma \ln{\left(\frac{F(\gamma)}{F(1-\gamma)}\right)} - \ln{\left(\frac{(1-e^{2\alpha})F(\gamma)}{\alpha(1-\gamma(1-e^\alpha))}\right)} -\frac{1}{\alpha}\coth{\frac{\alpha}{2}}\ln{\left(\frac{G(\gamma)G(1-\gamma) (1+G(0))}{1+G(1)}\right)}\\ 
        &+\frac{1}{\alpha}\tanh{\frac{\alpha}{2}}\left[\ln{\left(\frac{G(\gamma) F(\gamma)}{G(1-\gamma) F(1-\gamma)}\right)}+2\gamma \ln{\left(\frac{F(1-\gamma) G(1-\gamma)}{F(\gamma) G(\gamma)}\right)}\right] \\
        &+ \frac{1}{\alpha}\left[\frac{\pi^2}{6}-\Li_2(e^\alpha) + (1-2\gamma)\Li_2\left(\frac{\gamma+(1-\gamma)e^\alpha}{(1-e^\alpha)\gamma-1}\right) - (1-\gamma) \Li_2\left(\frac{e^\alpha (\gamma+(1-\gamma)e^\alpha)}{(1-e^\alpha)\gamma-1}\right) + \gamma \Li_2\left(\frac{\gamma \sinh{\alpha} -G(\gamma)}{\gamma \sinh{\alpha} +G(\gamma)}\right)\right]
    \end{split}
\end{equation}
with $F(\gamma) = \alpha (1-\gamma)-(1-2\gamma) \tanh{\frac{\alpha}{2}}$ and $G(\gamma)=1-\gamma +\gamma \cosh{\alpha}$.

\section{Optimal position and $\mathcal{I}_{\rm max}$ of clustered profile}\label{app:app2}

Far from thermodynamic equilibrium, where $\rho_L = 1$, $\rho_R = 0$ and $\gamma = 0$, in the limit $\alpha\uparrow\infty$ the density profile reduces to
\begin{equation}
    \label{eq:density_uniform_limit}
    \rho(\kappa) = (1-\frac{\kappa}{\kappa_0}) \theta(\kappa_0 - \kappa)\,,
\end{equation}
such that the average density is $\bar{\rho} = \kappa_0/2$. The PI simplifies greatly to
\begin{equation}
    \label{eq:positional_information_uniform_limit}
    \mathcal{I} = 1-\frac{\kappa_0 (1+\ln{\kappa_0}) + (2-\kappa_0) \ln{(2-\kappa_0)}}{\ln{4}}\,.
\end{equation}
Maximizing now the PI~\eqref{eq:positional_information_uniform_limit} with respect to $\kappa_0$, we find that the optimal cluster position is located at $\kappa_0 = 2/(1+e) = 1-\tanh{(1/2)}\approx 0.54$ with maximum PI $\mathcal{I}_{\rm max} = \log_2\left(\frac{1+\mathrm{e}}{\mathrm{e}}\right)\approx 0.452$. We numerically confirmed that this is the optimal value by assuming that $\gamma,\,\kappa_0$ and $\alpha$ are additional optimisation parameters and then computing the $\mathcal{I}-\Delta\rho-\Sigma$ Pareto front.

\section{WKB approximations}\label{app:app3}
Given the density profile resulting from applying the WKB approximation on the differential equation, \emph{i.e.,}
\begin{equation}
    \label{eq:solution_WKB_first_SM}
    \psi(\kappa) \sim \csch{\left(F(0,1)\right)} \left\{\psi_L \left(\frac{e_L}{e(\kappa)}\right)^\frac{1}{4}\sinh{\left(F(\kappa,1)\right)}
    +\psi_R \left(\frac{e_R}{e(\kappa)}\right)^\frac{1}{4}\sinh{\left(F(0,\kappa)\right)}\right\}\,,
\end{equation}
we can write the entropy production rate and reaction current as:
\begin{equation}
    \label{eq:entropy_production_WKB}
    \begin{split}
        \Sigma &\sim \frac{1}{\alpha}\left[\ln{\left(\frac{\gamma (1-\rho_L)}{\rho_L (1-\gamma)}\right)}\left(\psi_R(e_L\, e_R)^\frac{1}{4}\csch{F(0,1)} -\psi_L\sqrt{e_L} \coth{F(0,1)}-\frac{\psi_L}{4\alpha}\frac{e'_L}{e_L}\right)\right. \\
        &+\left.\ln{\left(\frac{\gamma (1-\rho_R)}{\rho_R (1-\gamma)}\right)}\left(\psi_L (e_L e_R)^\frac{1}{4} \csch{F(0,1)} -\psi_R \sqrt{e_R} \coth{F(0,1)}-\frac{\psi_R}{4\alpha}\frac{e'_R}{e_R}\right)\right]\,,
    \end{split}
\end{equation}
and 
\begin{equation}
    \label{eq:flux_WKB}
    \begin{split}
        \mathcal{J}\sim \alpha (\psi_L \sqrt{e_L} + \psi_R \sqrt{e_R})\coth{F(0,1)} - \alpha (\psi_L + \psi_R) (e_L e_R)^\frac{1}{4} \csch{F(0,1)} + \frac{1}{4}\left(\frac{\psi_L e'_L}{e_L} - \frac{\psi_R e'_R}{e_R}\right)\,,
    \end{split}
\end{equation}
with $F(s,t) = \alpha \int_s^t \mathrm{d}q \sqrt{e(q)}$, $e_L = e(0)$, $e'_L = e'(0)$, $e_R = e(1)$, $e'_R = e'(1)$ and $\psi_{L/R} = \rho_{L/R} - \gamma$.

\label{app:app2}